\documentclass{article}

\usepackage{fancyhdr}   
\usepackage{graphicx}   
\usepackage{geometry}   
\usepackage{PRIMEarxiv}
\usepackage{soul}    
\usepackage{xcolor}  
\usepackage[utf8]{inputenc} 
\usepackage[T1]{fontenc}    
\usepackage{hyperref}       
\usepackage{url}            
\usepackage{booktabs}       
\usepackage{amsfonts}       
\usepackage{nicefrac}       
\usepackage{microtype}      
\usepackage{lipsum}
\usepackage{fancyhdr}       
\usepackage{graphicx}       
\graphicspath{{media/}}     
\usepackage{booktabs} 
\usepackage{caption}  
\usepackage{array}    
\usepackage{siunitx} 
\usepackage{multirow}
\usepackage{booktabs}
\usepackage{tabularx} 
\usepackage{ragged2e} 
\usepackage{booktabs}
\usepackage{threeparttable}
\usepackage{tikz}
\usepackage{xcolor}
\usepackage{soul} 
\title{SecureBERT 2.0: Advanced Language Model for Cybersecurity Intelligence
}

\author{
  Ehsan Aghaei, Sarthak Jain, Prashanth Arun, Arjun Sambamoorthy \\
  Cisco AI\\
  San Jose, CA, USA\\
  \texttt{\{eaghaei, sarjain2, parun, asambamo\}@cisco.com} 
  \textit{}
}

\begin{document}
\maketitle

\begin{abstract}
Effective analysis of cybersecurity and threat intelligence data demands language models that can interpret specialized terminology, complex document structures, and the interdependence of natural language and source code. Encoder-only transformer architectures offer efficient and robust representations, supporting critical tasks such as semantic search, technical entity extraction, and semantic analysis—key to automated threat detection, incident triage, and vulnerability assessment. However, general-purpose language models typically lack the domain adaptation required for high precision in these contexts. We present SecureBERT 2.0, an enhanced encoder-only language model purpose-built for cybersecurity applications. Leveraging the ModernBERT architecture, SecureBERT 2.0 introduces improved long-context modeling and hierarchical encoding, enabling effective processing of extended and heterogeneous documents, including threat reports and source code artifacts. Pretrained on a domain-specific corpus over thirteen times larger than its predecessor, comprising more than 13 billion text tokens and 53 million code tokens from diverse real-world sources, SecureBERT 2.0 achieves state-of-the-art performance on multiple cybersecurity benchmarks. Experimental results demonstrate substantial improvements in semantic search for threat intelligence, semantic analysis, cybersecurity-specific named entity recognition, and automated vulnerability detection in code in cybersecurity domain.
\end{abstract}


\section{Introduction}
Cybersecurity is a field defined by its technical complexity, dynamic threat landscape, and the constant influx of vast, heterogeneous information, from threat intelligence feeds and research reports to raw source code and log files. Extracting actionable insight from this data requires not only domain expertise but also tools that can reliably interpret dense, jargon-heavy, and context-dependent language. Encoder-based, domain-specific language models are particularly well-suited for this task. Unlike general-purpose models, domain-specific architectures excel at creating deep, context-aware representations of input data, enabling precise understanding of relationships, terminology, and nuances critical to cybersecurity. When trained on security-specific corpora, these models can distinguish between benign and malicious patterns, accurately extract technical entities, and connect textual intelligence with code artifacts—capabilities that are foundational for automating threat detection, incident analysis, semantic similarity search, and vulnerability assessment in real-world environments.

Domain-specific models such as the original SecureBERT \cite{aghaei2022securebert} have already demonstrated the value of adapting language models to cybersecurity. However, the threat landscape has evolved: modern security workflows increasingly require a deeper understanding of technical text, structured documentation, and raw code, often in combination. This shift calls for a new generation of models that can capture richer context, longer-range dependencies, and subtle relationships specific to security data.

A key advancement in SecureBERT 2.0 is our adoption of the ModernBERT~\cite{warner2024smarter} architecture. Unlike older designs, ModernBERT addresses fundamental challenges in processing complex, hierarchical, and longer documents—commonplace in threat reports, log files, and source code. ModernBERT’s architectural improvements enable the model to retain and relate information over much longer sequences and across multiple modalities. This not only enhances the model’s ability to extract relevant details from sprawling security documents, but also improves its understanding of code structure and context—critical for tasks like vulnerability detection and technical entity extraction. In practice, this means more accurate results, better generalization to unseen threats, and a greater ability to reason about the connections between threat intelligence and operational code.

In addition to architectural advances, SecureBERT 2.0 benefits from a expanded pretraining dataset, thirteen times larger than that used for its predecessor. Our corpus unifies a vast array of cybersecurity text and code, drawn from real-world threat intelligence, technical blogs, incident reports, and open-source software repositories. This dual-modality approach is essential for today’s security environment, where critical information is often distributed across both prose and source code.

We validate these improvements across four downstream applications central to cyber operations: (1) document embedding with both bi-encoder and cross-encoder models for efficient semantic search and retrieval, (2) cybersecurity-focused named entity recognition, and (3) automated code vulnerability detection. SecureBERT 2.0 achieves new state-of-the-art performance in all evaluated tasks, setting fresh benchmarks for accuracy, coverage, and practical utility.

By bringing together a modern architecture tailored for complex, hybrid data and an unprecedented scale of domain-specific training, SecureBERT 2.0 delivers a new level of capability and reliability for cybersecurity practitioners and researchers working to stay ahead in a rapidly changing threat landscape.

\section{Pre-training}
\subsection{ModernBERT Architecture}
ModernBERT is a transformer-based language model designed specifically to overcome the challenges of cybersecurity data. Unlike traditional models, it introduces long-context handling through extended attention mechanisms, allowing it to process lengthy documents without losing critical information. It also features hierarchical encoding, enabling the model to understand both fine-grained and high-level structures common in technical text and source code. ModernBERT supports hybrid tokenization for both natural language and code, making it adept at multi-modal tasks. Efficient memory management allows it to scale to much larger datasets and longer sequences. These innovations make ModernBERT an ideal foundation for SecureBERT 2.0, enhancing its ability to model, analyze, and understand the complex, structured, and often lengthy data prevalent in cybersecurity.


\subsection{Dataset Specification}
To fully exploit the capabilities of ModernBERT and address the increasingly multifaceted nature of cybersecurity information, SecureBERT~2.0 is pretrained on a dataset that is substantially larger and more diverse than that used in the first version of SecureBERT. Our construction process emphasizes both coverage and quality, integrating a rich blend of textual cybersecurity corpora and source code repositories relevant to real-world security scenarios.

\paragraph{Data Sources.}
The pretraining corpus for SecureBERT 2.0 draws from several categories of cybersecurity-related datasets,
each contributing distinct linguistic and technical characteristics:

\begin{itemize}
    \item \textbf{Seed corpus}: High-quality curated security articles, reports, and technical blogs, providing foundational coverage of core cybersecurity concepts.
    \item \textbf{Large-scale web text}: A broad crawl of open web content filtered for relevance to cybersecurity, threat intelligence, and software security, contributing extensive language diversity.
    \item \textbf{Reasoning-focused data}: Security-oriented question answering and reasoning datasets, designed to enhance the model’s capabilities for inference and contextual understanding.
    \item \textbf{Instruction-tuning data}: Instructional and procedural texts relevant to cybersecurity tasks, supporting better generalization to task-oriented queries and automation.
    \item \textbf{Code vulnerability corpus}: A collection of annotated code samples focusing on software vulnerabilities, sourced from open-source projects and vulnerability databases.
    \item \textbf{Cybersecurity dialogue data}: Security-related conversational datasets capturing analyst workflows, technical Q\&A, and real-world interaction patterns.
    \item \textbf{Original baseline dataset}: The corpus used for pretraining the original SecureBERT model, included to maintain continuity and ensure coverage of established cybersecurity language.
\end{itemize}

\paragraph{Preprocessing and Quality Control.}

Given the heterogeneity of the sources, we applied a multi-stage pipeline to ensure high-quality, noise-reduced input:
\begin{enumerate}
\item \textbf{Deduplication and normalization}: Near-duplicate removal using MinHash and character-level similarity to reduce memorization risk, followed by Unicode normalization and consistent tokenization.
\item \textbf{Content filtering}: A hybrid approach combining keyword heuristics and lightweight cybersecurity classifiers to remove off-topic or low-signal data.
\item \textbf{Language balancing}: Dynamic re-weighting to prevent overrepresentation of high-volume sources (e.g., web crawls) and to maintain coverage of minority subdomains such as hardware security.
\end{enumerate}

\paragraph{Advanced Curriculum and Microannealing.}

To maximize the benefit of large-scale heterogeneous data, we incorporate microannealing, a fine-grained curriculum strategy that gradually adjusts sampling probabilities based on intermediate model perplexity:
\begin{itemize}
\item Early training favors high-quality, low-noise datasets (e.g., the seed corpus and curated vulnerability code), enabling the model to establish strong foundational representations.
\item Mid-stage sampling anneals toward more diverse sources, introducing challenging web text and reasoning datasets to expand linguistic and technical breadth.
\item Late training selectively revisits earlier high-quality sources to stabilize convergence and reduce catastrophic forgetting.
\end{itemize}
This microannealing pipeline allows SecureBERT 2.0 to progressively absorb complex security knowledge while maintaining stable optimization.


\paragraph{Dataset Statistics and Diversity.}
Table~\ref{tab:dataset_tokens} summarizes the composition of our pretraining corpus, detailing the number of code and text tokens contributed by each dataset. Notably, the combined dataset for SecureBERT~2.0 contains over \textbf{53.3 million code tokens} and more than \textbf{13.6 billion text tokens}, representing a thirteen-fold increase in data volume compared to the SecureBERT pretraining set.

This unprecedented scale ensures robust coverage of cybersecurity subdomains—including malware analysis, vulnerability research, incident response, and secure software development—while the integration of both code and text enables the model to bridge the gap between written threat intelligence and programmatic vulnerabilities.

\paragraph{Comparison with Previous SecureBERT Dataset.}
Compared to the original baseline model, which was pretrained on approximately 1.07 billion text tokens and no dedicated code corpus, the updated version leverages a dataset that is not only thirteen times larger in terms of text but also introduces tens of millions of code tokens. This expansion in both scale and modality allows the new model to achieve superior contextual understanding, improved generalization, and more effective context awareness capabilities that were limited in the first generation.

\begin{table}[h]
\centering
\renewcommand{\arraystretch}{1.2} 

\begin{tabular}{lrr}
\toprule
\textbf{Dataset Category} & \textbf{Code Tokens} & \textbf{Text Tokens} \\
\midrule
Seed corpus & 9,406,451 & 256,859,788 \\
Large-scale web text & 268,993 & 12,231,942,693 \\
Reasoning-focused data & -- & 3,229,293 \\
Instruction-tuning data & 61,590 & 2,336,218 \\
Code vulnerability corpus & 2,146,875 & -- \\
Cybersecurity dialogue data & 41,503,749 & 56,871,556 \\
Original baseline dataset & -- & 1,072,798,637 \\
\midrule
\textbf{Total} & 53,387,658 & 13,623,037,185 \\
\bottomrule
\end{tabular}
\caption{Token counts for code and text across categorized pretraining datasets.}
\label{tab:dataset_tokens}
\end{table}


Our model is trained on a comprehensive and carefully curated pretraining corpus that integrates publicly available security reports, vulnerability advisories, open-source cybersecurity white papers, technical books, and peer-reviewed research papers. This diverse collection enables the model to acquire rich contextual representations of cybersecurity knowledge spanning multiple subdomains. During pretraining, SecureBERT 2.0 captures fundamental security concepts such as network and system architecture, access control, authentication mechanisms, cryptographic protocols, threat intelligence, and overarching security principles. It also encodes detailed knowledge of software and system vulnerabilities, including common weakness patterns, exploitation techniques, secure configuration practices, patching strategies, and misconfiguration risks. Exposure to instructional and procedural materials allows the model to represent Tactics, Techniques, and Procedures (TTPs), security controls, compensating controls, operational workarounds, incident response workflows, and mitigation strategies. Training on source-code corpora further supports the learning of programming structures, secure coding practices, vulnerability patterns, and exploit mechanisms across multiple programming languages. In addition, SecureBERT 2.0 assimilates information from security dialogues, analyst reports, technical question-and-answer exchanges, and operational documents, enabling it to model analyst reasoning, threat scenarios, and policy or compliance considerations. The resulting embeddings provide a strong foundation for downstream tasks such as classification, information retrieval, code analysis, vulnerability assessment, and representation-based reasoning, positioning SecureBERT 2.0 as a versatile platform for advanced cybersecurity research and operational applications.

\subsection{Pretraining Objectives and Strategies}
The pretraining process for SecureBERT~2.0 was designed to maximize the model's ability to understand and represent both natural language and code relevant to cybersecurity. Our approach builds upon established masked language modeling techniques, with adaptations to account for the unique requirements of security-related corpora and code.

\paragraph{Pretraining Objectives.}
The primary objective for SecureBERT~2.0 pretraining is the masked language modeling (MLM) task, wherein a random subset of tokens in the input is dynamically masked, and the model is trained to predict the original tokens. This method encourages the model to develop contextualized representations that capture both the syntactic and semantic relationships within cybersecurity texts and code. For code-specific data, the MLM task is augmented by masking entire code identifiers and structural elements, helping the model to better learn code semantics and variable usage patterns which is an essential skill for tasks like vulnerability detection and source code analysis.

\paragraph{Training.}
Pretraining closely follows the procedures established in prior work but introduces several enhancements:
\begin{itemize}
    \item \textbf{Architecture:} We utilize ModernBERT, a contemporary update to previous BERT variants, chosen for its improved efficiency and accuracy in modeling long and complex sequences. 
    \item \textbf{Data Pipeline:} A robust data cleaning pipeline filters, deduplicates, and tokenizes data prior to training, ensuring high-quality and relevant input from both text and code sources.
    \item \textbf{Pretraining Duration:} The model was trained for 20 epochs over the aggregated dataset, allowing thorough exposure and representation learning across diverse cybersecurity data.
    \item \textbf{Token Length:} All input sequences were standardized to a length of 1024 tokens, providing the model with substantial context for both documents and code samples.
    \item \textbf{Optimizer:} We employed the AdamW optimizer, which is well-suited for large-scale transformer models.
\end{itemize}

\paragraph{Hyperparameters and Infrastructure.}
Table \ref{tab:training_hyperparams} reflects the pre-training hyperparameters. Training was conducted on a multi-GPU infrastructure, enabling efficient large-batch learning and timely convergence.
\begin{table}[h!]
\centering
\begin{tabular}{|l|l|l|}
\hline
\textbf{Parameter} & \textbf{Value} & \textbf{Notes} \\ \hline
\texttt{batch\_size}    & 16   & Per GPU (using 8 GPUs in parallel) \\ \hline
\texttt{learning\_rate} & 5e-5 & -- \\ \hline
\texttt{weight\_decay}  & 0.01 & -- \\ \hline
\texttt{mlm\_prob}      & 0.10 & Probability of a token being masked \\ \hline
\end{tabular}
\caption{Pre-training hyperparameters.}
\label{tab:training_hyperparams}
\end{table}

\section{MLM Performance Evaluation}
In our evaluation, we masked only verbs or nouns in each sentence, reflecting the domain-specific observation that verbs (representing actions) and nouns (denoting objects) carry the most semantic significance in cybersecurity for describing threats, techniques, and procedures. For source code data, we applied targeted masking to specific code elements, including function names, variable identifiers, and operators, while preserving structural tokens such as keywords and delimiters. The test set consists of 12,721 sentences with a masked noun (covering 2,213 unique nouns), 4,620 sentences with a masked verb (covering 888 unique verbs), and 8,540 code snippets with masked code tokens, covering 3,102 unique identifiers and 1,245 operators. All test cases were strictly excluded from the pretraining corpus to ensure that evaluation reflects the model’s generalization to unseen data.

\paragraph{Quantitative Results.}
Table~\ref{tab:mlm_results} presents the top-$n$ accuracy scores for object (noun, e.g., \textit{malware}, \textit{firewall}, \textit{phishing}), verb (e.g., \textit{encrypt}, \textit{bypass}, \textit{exploit}), and code-specific (e.g., \texttt{printf()}, \texttt{strcpy()}, \texttt{chmod}, \texttt{ls -la}, \texttt{}) MLM tasks. SecureBERT~2.0 demonstrates strong performance on both masked objects and actions, significantly outperforming general-domain models in predicting contextually relevant cybersecurity terms. Notably, the model achieves top-1 accuracies of 56.20\% (objects) and 45.02\% (verbs), and top-5 accuracies exceeding 82\% and 74\%, respectively. For code-based MLM, the model also achieves solid performance, with top-1 accuracy at 39.27\% and top-5 at 55.41\%. 

\begin{table}[h]
    \centering
    \renewcommand{\arraystretch}{1.1}
    \begin{tabular}{lccc}
        \toprule
        \textbf{Top-$n$} & \textbf{Objects (Nouns)} & \textbf{Verbs (Actions)} & \textbf{Code\_MLM} \\
        \midrule
        1  & 56.20\% & 45.02\% & 39.27\% \\
        2  & 69.73\% & 60.00\% & 46.90\% \\
        3  & 75.85\% & 66.68\% & 50.87\% \\
        4  & 80.01\% & 71.56\% & 53.36\% \\
        5  & 82.72\% & 74.12\% & 55.41\% \\
        10 & 88.80\% & 81.64\% & 60.03\% \\
        \bottomrule
    \end{tabular}
    \caption{MLM accuracy (\%) for predicting masked objects (nouns), verbs (actions), and code tokens in cybersecurity context.}
    \label{tab:mlm_results}
\end{table}

\paragraph{Qualitative Analysis.}
SecureBERT~2.0 demonstrates a markedly stronger grasp of domain-specific semantics compared to original SecureBERT and other baseline models. In challenging cases where the masked word corresponds to specialized cybersecurity terms (e.g., ``reconnaissance,'' ``hijacking,'' or ``DDoS''), SecureBERT~2.0 consistently produces correct predictions, whereas generic models such as ModernBERT often fail to generate contextually relevant completions. This superior contextual understanding is essential for downstream tasks such as incident analysis, threat detection, and context-aware cyber threat intelligence.
\begin{figure}[h]
    \centering
    \includegraphics[width=1\linewidth]{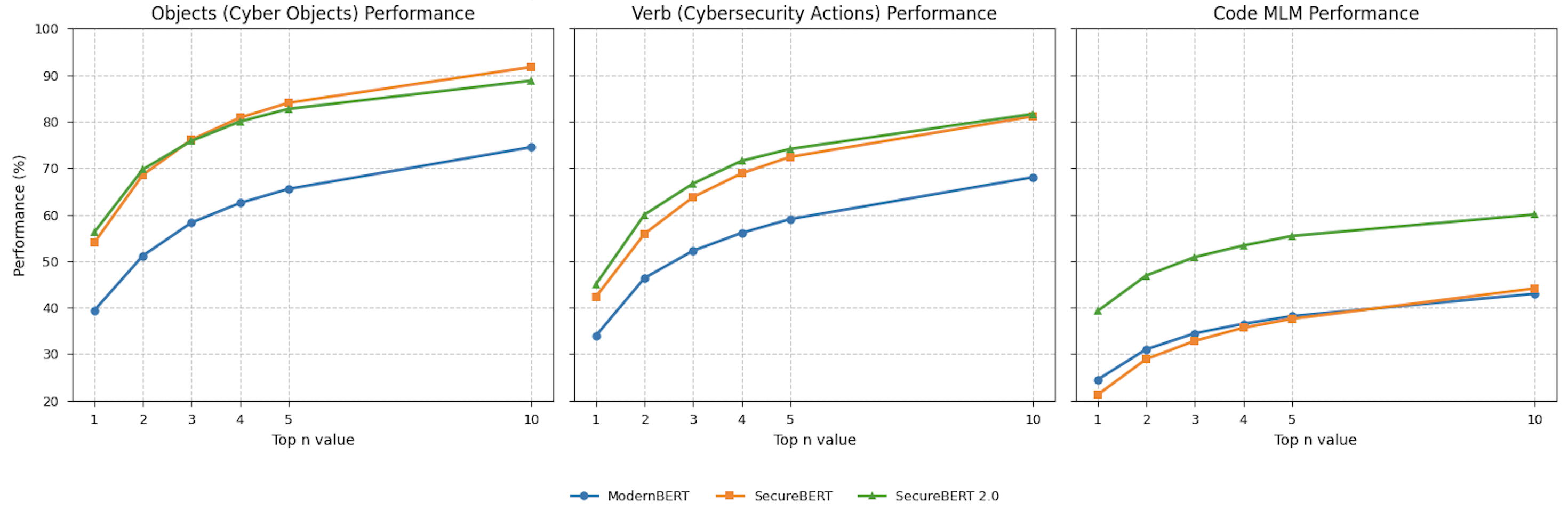}
    \caption{Comparison of MLM performance in predicting objects, verbs, and code tokens.}
    \label{fig:mlm_comparison}
\end{figure}

Fig.~\ref{fig:mlm_comparison} demonstrates that SecureBERT~2.0 outperforms both the original SecureBERT and ModernBERT in the MLM evaluation, particularly in code understanding. Although its Top-10 performance is slightly lower than that of the original SecureBERT, this difference reflects design and training choices rather than reduced model quality. The original SecureBERT was trained on a smaller dataset, whereas SecureBERT~2.0 has been trained on a larger and more diverse corpus, encompassing both natural language and source code. These factors produce richer, domain-specific embeddings and a sharper probability distribution that concentrates on the most contextually relevant tokens. As a result, Top-1 through Top-5 predictions, which are more indicative of the model's precision in token prediction, show clearer improvements, reflecting higher-quality contextual representations.

SecureBERT~2.0 also outperforms original SecureBERT and all other state-of-the-art models in the code-level MLM task. While MLM is not a standard benchmarking metric for evaluating a language model, it provides a useful proxy for assessing the overall understanding of model's context awareness. High performance on code MLM indicates that the model has learned meaningful representations of programming constructs, cybersecurity terminology, and their contextual interactions. This capability is especially relevant for tasks such as vulnerability analysis, code review, and context-aware threat intelligence, demonstrating that SecureBERT~2.0 captures richer domain knowledge beyond what similar models can reveal.




\section{Downstream Tasks and Application Design}
To comprehensively evaluate the capabilities of SecureBERT~2.0, we applied the model to a diverse set of downstream tasks representative of real-world cybersecurity challenges. These tasks were selected to highlight the model's ability to understand and reason over both natural language and code, as well as its effectiveness in practical applications that are critical to security operations. In the following subsections, we detail the design, datasets, and evaluation protocols for each application area, including document embedding, named entity recognition, and code vulnerability detection.

\subsection{Document Embedding Model}
Document embeddings have emerged as a crucial component in the cybersecurity domain, enabling efficient processing and utilization of vast and complex textual data. These embeddings facilitate a variety of applications including Retrieval-Augmented Generation (RAG), information retrieval, document ranking and re-ranking, and semantic search. In RAG frameworks, embeddings assist in retrieving contextually relevant documents that enhance the quality and accuracy of generated outputs. Similarly, embedding-based ranking systems improve the prioritization of security advisories, vulnerability reports, and incident descriptions, enabling analysts to focus on the most pertinent information. Embedding-driven document search further enables semantic matching beyond keyword-based methods, improving retrieval precision in tasks such as threat hunting, compliance checking, and knowledge management.

To leverage these applications effectively, it is important to employ both bi-encoder and cross-encoder architectures, each offering distinct advantages. Bi-encoders independently encode queries and documents into a shared vector space, allowing for efficient approximate nearest neighbor search over large datasets. This scalability makes bi-encoders suitable for initial retrieval stages where rapid candidate selection is essential. However, they inherently lack the capacity to capture detailed interactions between query and document tokens, which may limit ranking accuracy. Cross-encoders address this limitation by jointly encoding the query-document pair, thereby modeling fine-grained contextual dependencies. Although cross-encoders achieve superior performance in ranking and re-ranking, their higher computational cost restricts their use to smaller candidate sets. By combining both architectures in a two-stage pipeline, we obtain a balance between retrieval speed and ranking precision, optimizing performance for real-world cybersecurity scenarios.

We fine-tuned our document embedding models, including both bi-encoder and cross-encoder architectures, using multiple cybersecurity-specific datasets. The following categories describe the data used for fine-tuning and evaluation.

\paragraph{Cybersecurity QA corpus:}  
This dataset contains approximately 43,000 records composed of question–answer pairs, incident reports, and domain knowledge snippets within cybersecurity. It spans diverse subdomains such as network security, malware analysis, cryptography, and cloud security. The dataset features varied text lengths and rich technical terminology sourced from open threat intelligence and security documentation, making it well-suited for fine-tuning embedding models aimed at both precise retrieval and broad contextual understanding. 

\paragraph{Security governance QA corpus:}  
This dataset includes about 60,000 curated question–answer pairs focusing on governance, vulnerability management, compliance, and exploit analysis. The data emphasizes concise, expert-validated answers across diverse question types, enhancing the model's ability to generalize semantic representations in the security domain.

\paragraph{Cybersecurity instruction–response corpus:}  
This dataset contains roughly 25,000 instruction–response pairs designed to train models on instruction following and contextual reasoning in cybersecurity. Instructions such as “Describe mitigation techniques for cross-site scripting” are paired with detailed, domain-specific technical responses. This corpus supports fine-tuning for improved reranking with cross-encoders and enhanced semantic search with bi-encoders.

\paragraph{Cybersecurity rules corpus (evaluation):}  
For evaluation, we employ a dataset comprising approximately 5,000 structured cybersecurity policies, guidelines, and best practice statements. This includes normative security knowledge represented in natural language and domain-specific terminology. This dataset provides a rigorous benchmark to assess the retrieval quality of embedding models in matching queries against security standards and compliance rules.

\vspace{0.5cm}

\begin{table}[h]
\centering
\renewcommand{\arraystretch}{1.2}
\begin{tabular}{@{}lr@{}}
\toprule
\textbf{Dataset Category} & \textbf{Number of Records}  \\
\midrule
Cybersecurity QA corpus & 43,000  \\
Security governance QA corpus & 60,000 \\
Cybersecurity instruction–response corpus & 25,000 \\
Cybersecurity rules corpus (evaluation) & 5,000  \\
\bottomrule
\end{tabular}
\caption{Summary of categorized cybersecurity datasets used for fine-tuning and evaluation.}
\label{tab:dataset_summary}
\end{table}

Our training strategy involved further fine-tuning SecureBERT 2.0 on cybersecurity datasets using both bi-encoder and cross-encoder objectives. The bi-encoder models were optimized with a contrastive learning framework that encourages high similarity between semantically related text pairs while reducing similarity for negative pairs, including hard negatives dynamically mined during training. In parallel, the cross-encoder models were fine-tuned with a binary classification objective to distinguish relevant from irrelevant pairs. Model optimization employed the AdamW optimizer with a learning rate schedule incorporating an initial warm-up phase followed by linear decay. Training was performed on NVIDIA GPUs with mixed-precision computation to improve efficiency and reduce memory usage, and early stopping based on validation performance was applied to prevent overfitting. Hyperparameters were systematically tuned to achieve robust performance across datasets of varying sizes and characteristics, yielding embeddings that generalize effectively to diverse downstream cybersecurity tasks.

\subsection{Named Entity Recognition (NER)}

Named Entity Recognition (NER) plays a pivotal role in cybersecurity by enabling the automatic extraction and classification of critical entities from unstructured text data, such as threat intelligence reports, security advisories, and incident logs. Identifying entities like indicators of compromise (IOCs), malware names, affected systems, organizations, and vulnerabilities facilitates the automation of threat analysis, enhances situational awareness, and supports rapid response to security incidents. NER models aid in building structured knowledge bases, enriching threat intelligence platforms, and improving the efficiency of security operations by transforming raw textual data into actionable insights.

To evaluate the performance of SecureBERT 2.0 on NER, we utilized a selection of prominent cybersecurity-specific datasets. These datasets are crucial for assessing the model's ability to identify and classify domain-specific entities within unstructured text, a fundamental task in cyber threat intelligence.

For the NER task, we utilized several publicly available benchmark corpus providing annotated cybersecurity-specific entities. 
This dataset serves as a critical resource for training and evaluating domain-adapted NER models. 
Table~\ref{tab:ner_corpus} summarizes its key characteristics, including purpose, data sources, annotation methodology, 
defined entity types, and overall size.

\begin{table}[h!]
\centering
\caption{Overview of the benchmark cybersecurity NER corpus}
\label{tab:ner_corpus}
\begin{tabular}{p{4cm} p{10cm}}
\toprule
\textbf{Aspect} & \textbf{Description} \\
\midrule
\textbf{Purpose} & A manually annotated benchmark dataset for extracting cybersecurity concepts from unstructured threat intelligence reports; designed as a foundational resource for training and evaluating NER models. \\
\midrule
\textbf{Data Source} & Derived primarily from high-quality, noise-free threat intelligence reports, with emphasis on malware analysis. \\
\midrule
\textbf{Annotation Methodology} & Fully hand-labeled by domain experts to ensure accuracy, consistency, and contextual relevance. \\
\midrule
\textbf{Entity Types} & Defines five entity categories: \textit{Malware, Indicator, System, Organization, Vulnerability}. \\
\midrule
\textbf{Size} & Contains 3.4k samples in the combined train set and 717 rows for testing. \\
\bottomrule
\end{tabular}
\end{table}
 The model was optimized with a token-wise Cross Entropy loss and trained using the AdamW optimizer with a linear learning rate scheduler. To ensure stability, gradient clipping was applied with a maximum norm of 1.0.
Key hyperparameters included a maximum sequence length of 1024, a per-GPU batch size of 8, a learning rate of 1e-5, and a weight decay of 0.001, trained for 20 epochs.

\subsection{Code Vulnerability Detection}
Code Vulnerability Detection plays a critical role in cybersecurity by enabling the automatic identification and flagging of security weaknesses within source code. This process is essential for proactively preventing potential exploits, reducing the attack surface of applications, and ensuring the overall integrity and confidentiality of software systems. Identifying vulnerabilities such as SQL injection, cross-site scripting (XSS), buffer overflows, insecure deserialization, and misconfigurations allows developers and security teams to remediate flaws early in the software development lifecycle (SDLC). Effective code vulnerability detection aids in automating security testing, minimizing remediation costs, and fostering a robust security posture by transforming raw code into actionable insights for secure development. To evaluate the performance of SecureBERT 2.0 on Code Vulnerability Detection, we utilized a selection of prominent cybersecurity-specific codebases and vulnerability datasets. These datasets are crucial for assessing the model’s ability to pinpoint and classify domain-specific flaws within program code, a fundamental task in secure software engineering.

\begin{table}[h!]
\centering
\caption{Overview of the code vulnerability detection dataset}
\label{tab:vuln_dataset}
\begin{tabular}{p{4cm} p{10cm}}
\toprule
\textbf{Aspect} & \textbf{Description} \\
\midrule
\textbf{Purpose} & A benchmark corpus for vulnerability detection in source code, designed to support the training and evaluation of models for identifying security flaws and aiding program repair research. \\
\midrule
\textbf{Data Source} & Curated from open-source C/C++ software repositories, with both vulnerable and non-vulnerable code functions extracted for balanced coverage. \\
\midrule
\textbf{Annotation Methodology} & Vulnerability labels derived from commits associated with security fixes, combining commit metadata, static analysis tools, and manual verification for accuracy. Each function is labeled as either vulnerable or non-vulnerable. \\
\midrule
\textbf{Label} & Binary vulnerability status of each code function (\textit{True} or \textit{False}). \\
\midrule
\textbf{Size} & Approximately 21.9k samples in the training set and 2.7k samples in the test set. \\
\bottomrule
\end{tabular}
\end{table}

Our training strategy involved fine-tuning the SecureBERT 2.0 model for binary sequence classification on this code vulnerability dataset. The model was fine-tuned using a sequence classification loss. We utilized the AdamW optimizer with a linear learning rate scheduler. Gradient clipping with a maximum norm of 1.0 was applied to the model parameters. Key hyperparameters included a maximum sequence length of 1024, a per-GPU batch size of 8, a learning rate of 1e-5, and a weight decay of 0.01. The training ran for 10 epochs, and the model weights were initialized from a masked language modeling (MLM) pre-existing checkpoint.


    
    
    
    

\section{Experimental Results}
\subsection{Document Embedding Model}
We evaluated \textbf{SecureBERT~2.0}’s ability to generate meaningful and discriminative embeddings using both bi-encoder and cross-encoder architectures. Performance was benchmarked against several strong baselines, including \textbf{AttackBERT}\cite{abdeen2023smet}, \textbf{all-MiniLM-L6-v2}\cite{wang2020minilm}, and other high-performing models from the Hugging Face model zoo. 

Evaluation metrics include mean Average Precision (\textbf{mAP}), Recall at 1 (\textbf{R@1}), Mean Reciprocal Rank at 10 (\textbf{MRR@10}), and Normalized Discounted Cumulative Gain at 10 (\textbf{NDCG@10}), providing a comprehensive assessment of retrieval and ranking quality.

\subsubsection{Cross-Encoder Results}
The cross-encoder jointly encodes the query and document, allowing fine-grained interaction and semantic matching. As shown in Table~\ref{tab:docembed_cross}, \textbf{SecureBERT~2.0} achieves a remarkable \textbf{mAP of 0.955} and \textbf{R@1 of 0.948}, outperforming all baselines across every metric. In particular, it surpasses the well-established \texttt{ms-marco-MiniLM-L6-v2} and \texttt{TinyBERT-L2-v2} models by over 3–5 percentage points on mAP and R@1. The improvements in \textbf{NDCG@10 (0.986)} and \textbf{MRR@10 (0.983)} further highlight its superior ranking and semantic precision, demonstrating exceptional understanding of cybersecurity-related text.

\begin{table}[h]
\centering
\caption{Document Embedding Performance (Cross-Encoder)}
\label{tab:docembed_cross}
\renewcommand{\arraystretch}{1.2}
\begin{tabular}{lcccc}
\toprule
\textbf{Model} & \textbf{mAP} & \textbf{R@1} & \textbf{NDCG@10} & \textbf{MRR@10} \\
\midrule
BAAI/bge-reranker-large & 0.741 & 0.741 & 0.894 & 0.864 \\
bge-reranker-base & 0.781 & 0.709 & 0.896 & 0.868 \\
cross-encoder/ms-marco-MiniLM-L6 & 0.882 & 0.859 & 0.950 & 0.937 \\
cross-encoder/ms-marco-MiniLM-L4-v2& 0.904 & 0.863 & 0.964 & 0.956 \\
cross-encoder/ms-marco-TinyBERT-L2-v2 & 0.920 & 0.849 & 0.964 & 0.955 \\
\textbf{SecureBERT 2.0} & \textbf{0.955} & \textbf{0.948} & \textbf{0.986} & \textbf{0.983} \\
\bottomrule
\end{tabular}
\end{table}

\subsubsection{Bi-Encoder Results}
The bi-encoder configuration encodes queries and documents independently, providing higher scalability and lower inference latency—ideal for large-scale retrieval and real-time RAG pipelines. Table~\ref{tab:docembed_bi} shows that \textbf{SecureBERT~2.0} attains an exceptional \textbf{mAP of 0.951} and \textbf{R@1 of 0.984}, outperforming all baselines, including domain-specific and large transformer models. Notably, despite the typical expressivity gap between bi- and cross-encoders, SecureBERT~2.0 closes this gap almost entirely, achieving near-identical ranking performance while being significantly faster in retrieval operations.
\begin{table}[h]
\centering
\caption{Document Embedding Performance (Bi-Encoder)}
\label{tab:docembed_bi}
\renewcommand{\arraystretch}{1.2}
\begin{tabular}{lccc}
\toprule
\textbf{Model} & \textbf{mAP} & \textbf{R@1} & \textbf{MRR@10} \\
\midrule
krlng/sts-GBERT-bi-encoder & 0.672 & 0.806 & 0.852 \\
sentence-transformers/all-MiniLM-L6-v2 & 0.898 & 0.917 & 0.940 \\
basel/ATTACK-BERT & 0.906 & 0.918 & 0.944 \\
sentence-transformers/all-MiniLM-L12-v2 & 0.912 & 0.924 & 0.945 \\
\textbf{SecureBERT 2.0} & \textbf{0.951} & \textbf{0.984} & \textbf{0.989} \\
\bottomrule
\end{tabular}
\end{table}

\subsubsection{Analysis and Discussion}
\textbf{SecureBERT~2.0} demonstrates state-of-the-art retrieval and ranking performance across both encoder architectures. Several key observations emerge:
\begin{itemize}
    \item \textbf{Superior Domain Adaptation:} The model’s advantage over AttackBERT and MiniLM variants reinforces the benefits of pretraining on a large-scale, diverse cybersecurity corpus. SecureBERT~2.0 captures nuanced terminology and contextual dependencies unique to the cybersecurity domain.
    \item \textbf{Cross-Architecture Consistency:} Achieving near-parity between cross- and bi-encoder results is rare, underscoring the embedding robustness and semantic depth of SecureBERT~2.0. This flexibility enables both accuracy-sensitive and latency-sensitive deployment scenarios.
    \item \textbf{Scalability and Efficiency:} The bi-encoder’s superior R@1 and mAP make SecureBERT~2.0 ideal for high-throughput applications such as retrieval-augmented generation (RAG), automated threat intelligence systems, and analyst-support tools where fast, accurate retrieval is essential.
    \item \textbf{Ranking Precision:} The consistently high MRR and NDCG values indicate that SecureBERT~2.0 not only retrieves relevant results but also ranks them effectively, ensuring downstream models and analysts receive the most contextually relevant information first.
\end{itemize}

\subsection{Named Entity Recognition}
This model enables the systematic identification of key entities such as vulnerabilities, malware, systems, indicators, and organizations. In this section, we present an initial showcase of SecureBERT 2.0 on these five core entity types as a general demonstration of its capability for cybersecurity-specific information extraction. Performance is assessed through a comparative analysis against established baselines using standard metrics of precision, recall, and F1-score. While this evaluation focuses on a controlled experimental setting, the results provide a strong indication of the model’s capacity to support real-world applications, where entity recognition underpins downstream tasks such as threat intelligence enrichment, attack graph construction, and automated defensive response. Future extensions of this work aim to incorporate additional entity categories, including tactics, techniques, and procedures, further enhancing the model’s applicability to comprehensive cybersecurity analysis.

\subsubsection{Quantitative Results}
We evaluate the performance of all experiments using \textbf{precision}, \textbf{recall}, and \textbf{F1-score}. Precision quantifies the proportion of predicted entities that are correct, recall measures the proportion of true entities that are successfully identified, and the F1-score represents the harmonic mean of precision and recall. Table~\ref{tab:ner_results} presents the results obtained across different training regimens and baseline models.

\begin{table}[h]
    \centering
    \renewcommand{\arraystretch}{1.2}
    \caption{NER Results on CyNER Test Set}
    \label{tab:ner_results}
    \begin{tabular}{lccc}
        \toprule
        \textbf{Model / Training Regimen} & \textbf{F1-score} & \textbf{Recall} & \textbf{Precision} \\
        \midrule
        CyBERT (fine-tuned on NER) & 0.3509 & 0.2810 & 0.4671 \\
        SecureBERT ((fine-tuned on NER) & 0.7340 & 0.7588 & 0.7172 \\
        \textbf{SecureBERT 2.0} & \textbf{0.945} & \textbf{0.965 }& \textbf{0.927 }\\
        \bottomrule
    \end{tabular}
\end{table}

\subsubsection{Analysis and Discussion}
The results in Table~\ref{tab:ner_results} clearly highlight the performance gap between SecureBERT 2.0 and prior baselines. CyBERT, which represents an earlier effort toward domain-specific adaptation, achieved relatively low F1 (0.3509), with recall lagging significantly (0.2810), suggesting difficulty in consistently detecting relevant entities. SecureBERT substantially improved this baseline, reaching an F1 of 0.7340 and balanced precision and recall, reflecting the value of cybersecurity-oriented pretraining. 
Our proposed model demonstrates a further and substantial leap in performance, achieving an F1-score of 0.945 with near-perfect recall (0.965) and very high precision (0.927). This indicates that the model is not only effective at identifying the majority of true entities but also does so with a low rate of false positives. The strong recall performance is especially valuable in cybersecurity contexts, where missed entities (e.g., overlooked malware names or vulnerabilities) could undermine downstream tasks such as automated threat correlation and situational awareness.

The high precision ensures that extracted entities are trustworthy enough to feed into automated pipelines, reducing the burden of post-processing or human verification. These gains can be attributed to a combination of architectural optimizations, domain-specific pretraining, and fine-tuning strategies tailored to cybersecurity text. Importantly, the improvements suggest that the model generalizes well beyond training data, capturing linguistic patterns and contextual cues specific to technical reporting and threat intelligence narratives.

Nonetheless, while the results are highly promising, they should be interpreted with awareness of the controlled experimental setup. Real-world cyber threat intelligence data often contains noisy, incomplete, or adversarially manipulated information, which may impact performance. Further evaluation on diverse, heterogeneous datasets is necessary to fully establish robustness.

\subsubsection{Qualitative Examples.}
To better illustrate the effectiveness of our approach, we present qualitative examples highlighting extracted entities across different cybersecurity contexts. These examples demonstrate how the model identifies and classifies key concepts such as malware, vulnerabilities, indicators, systems, and organizations within natural text, providing structured insights that are critical for downstream analysis.

\definecolor{MalwareColor}{RGB}{255,102,102}        
\definecolor{IndicatorColor}{RGB}{102,178,255}     
\definecolor{SystemColor}{RGB}{102,255,178}        
\definecolor{VulnerabilityColor}{RGB}{255,178,102}
\definecolor{OrganizationColor}{RGB}{178,102,255}  

\newcommand{\entity}[3]{%
  \tikz[baseline=(word.base)]{
    \node[fill=#1!30, rounded corners=2pt, inner sep=1pt] (word) {#3};
    \node[above=1mm, font=\tiny, anchor=south, fill=#1!60, rounded corners=1pt] at (word.north) {#2};
  }%
}

\begin{itemize}
    \item \textit{"The \entity{MalwareColor}{Malware}{Emotet} campaign exploited \entity{IndicatorColor}{Indicator}{malicious email attachments} to target \entity{OrganizationColor}{Organization}{financial institutions}."}
    \item \textit{"A \entity{VulnerabilityColor}{Vulnerability}{buffer overflow} in the \entity{SystemColor}{System}{Windows Server} enabled remote code execution."}
    \item \textit{"Observed indicators included \entity{IndicatorColor}{Indicator}{IP address 192.0.2.1} and \entity{IndicatorColor}{Indicator}{MD5 hash e99a18c...} during the attack."}
\end{itemize}

\subsection{Code Vulnerability Detection}
Code vulnerability detection plays a central role in strengthening software security, enabling proactive identification of weaknesses before they can be exploited. Given a code snippet, this model returns a binary prediction on whether it contains any vulnerability or not.

We report precision, recall, and accuracy for all experiments. Precision measures the proportion of predicted entities
that are correct, recall measures the proportion of true entities that are recovered, and accuracy is the percentage of total examples answered correctly.

Table \ref{tab:code_vuln_accuracy} compares our model with existing approaches for detecting code vulnerabilities. Our model achieves the highest accuracy of 0.655, showing it can correctly classify code snippets more consistently than the baselines. CodeBERT, while precise (0.821), misses a large number of actual vulnerabilities due to its low recall (0.241). CyBERT, on the other hand, catches all vulnerabilities (recall 1.0) but produces many false positives because of its lower precision (0.459). Overall, our model strikes a better balance between precision (0.631) and recall (0.602), resulting in a higher F1 score (0.616) and demonstrating reliable detection without overwhelming false alarms.

\begin{table}[h!]
    \centering
    \caption{Model Accuracy Comparison}
    \label{tab:code_vuln_accuracy}
    \begin{tabular}{lcccc} 
        \toprule
        \textbf{Model} & \textbf{Accuracy} & \textbf{F1 Score} & \textbf{Recall} & \textbf{Precision}\\
        \midrule

        CodeBERT~\cite{feng2020codebert} & 0.627 & 0.3722 & 0.2406 & 0.8207 \\
        CyBERT~\cite{ranade2021CyBERT2} & 0.459 & 0.6295 & 1.0000 & 0.4594 \\
        \textbf{SecureBERT 2.0}& \textbf{0.655} & \textbf{0.616} & \textbf{0.602} & \textbf{0.630} \\
        \bottomrule
    \end{tabular}
\end{table}

\section{Related Works}
\subsection{Domain-Specific Pretrained Models for Cybersecurity}

The original SecureBERT \cite{aghaei2022securebert} marked a pivotal step in adapting large language models (LLMs) for cybersecurity. Pretrained on a specialized corpus of threat reports, vulnerability descriptions, and technical blogs, SecureBERT demonstrated that domain-specific pretraining could yield significant improvements in core cybersecurity tasks, including entity extraction, document understanding, and contextual reasoning about threats. By showing that security-oriented corpora could bridge the gap left by general-purpose models such as BERT \cite{Devlin2019BERTPO} and RoBERTa \cite{Liu2019RoBERTaAR}, SecureBERT provided the first strong evidence that LLMs can meaningfully support cyber defense and threat intelligence workflows. However, its training was constrained by a relatively modest dataset and limited to textual modalities, restricting its ability to generalize across the full spectrum of cybersecurity data.

Prior to SecureBERT, other efforts explored domain adaptation of transformers for cybersecurity. CyBERT focused on fine-tuning BERT for named entity recognition (NER) and threat extraction from security advisories and reports. Similarly, AttackBERT \cite{abdeen2023smet} specialized in adversarial behavior descriptions, supporting semantic retrieval of attack narratives. While these models demonstrated improvements over general-purpose LMs, they were limited in coverage and generalization compared to SecureBERT.

Following SecureBERT, several models have emerged to address specialized aspects of cyber defense. CyberBERT \cite{paul2022cyberbert} targeted textual advisories and structured threat reports, VulBERTa \cite{hanif2022vulberta} emphasized vulnerability detection in source code, and CySecBERT \cite{bayer2024cysecbert} adopted a generative pretraining approach, enabling both threat text generation and security question answering. These works collectively validate the benefits of domain-specific pretraining, but many are still constrained by small datasets, focus on single modalities, or lack architectures capable of handling long-context inputs from multi-source cybersecurity data.

\subsection{Models for Code Understanding and Vulnerability Detection}

In parallel, research has focused on pretrained models for programming languages and code reasoning. CodeBERT \cite{feng2020codebert} and GraphCodeBERT \cite{guo2020graphcodebert} introduced joint pretraining on source code and natural language, supporting tasks such as function documentation, code search, and defect detection. CodeT5 \cite{wang2021codet5} and PLBART \cite{ahmad2021unified} further advanced this line by adopting encoder–decoder architectures for program repair and code generation.

In the context of security, datasets such as Devign \cite{zhou2019devign} and DetectVul have enabled benchmarking of vulnerability detection, while models such as VulBERTa \cite{hanif2022vulberta} explicitly target security flaws in C/C++ code. Despite these advances, most existing code-pretrained models are not explicitly designed for cybersecurity contexts, limiting their effectiveness in capturing threat semantics or reasoning across heterogeneous data sources such as reports and source code.

\subsection{Hybrid and Multimodal Security Models}

As cybersecurity knowledge spans both natural language text (e.g., advisories, threat reports) and source code (e.g., vulnerable functions, patches), there has been increasing interest in hybrid models. Works such as UniXcoder \cite{guo2022unixcoder} and PolyCoder \cite{xu2022polycoder} have demonstrated the utility of models that can simultaneously process multiple modalities of technical information. In security, recent explorations such as CySecBERT and SecureBERT~2.0 suggest that combining threat intelligence with code corpora can enhance reasoning about vulnerabilities and exploits. However, these approaches are often limited by smaller training sets and the lack of long-context architectures necessary for handling lengthy technical reports or multi-file codebases.

\subsection{Recent Generative Models in Cybersecurity}

Recent advancements in generative artificial intelligence (GenAI) have led to the development of specialized models tailored for cybersecurity applications. These models extend the capabilities of traditional large language models (LLMs) by enabling tasks such as threat report synthesis, exploit scenario generation, and automated vulnerability summarization.

For instance, Colibri~\cite{colibri_8b}, developed by CyberNative-AI, is a conversational AI fine-tuned for cybersecurity tasks. It has been evaluated using the FARR Flow reasoning method to assess its ability to guide penetration testers through testing processes. Additionally, the PurpleLlama CyberSecEval framework has been employed to measure Colibri's comprehension of the MITRE ATT\&CK knowledge base.

Another notable model is CyberBase~\cite{cyberbase_13b}, also from CyberNative, which serves as a base model for cybersecurity. Built upon the Vicuna architecture, CyberBase has been fine-tuned on cybersecurity-related data, specifically GitHub cybersecurity READMEs. It is designed for future fine-tuning and is not recommended for standalone use.

ZySec-7B~\cite{zysec_7b} is an open-source AI model for cybersecurity, trained across over 30 domains, including attack surface threats, cloud security, and compliance frameworks like PCI DSS and ISO/IEC 27001. It offers on-demand expert guidance and integrates AI into cybersecurity operations, enhancing strategic decision-making and risk management.

DeepHat-V1-7B~\cite{deephat_1_7b} is a model series designed for both offensive and defensive cybersecurity tasks. It provides capabilities for threat analysis, vulnerability assessment, and incident response, supporting cybersecurity professionals in various operational contexts.

Foundation-Sec-8B~\cite{foundation_sec_8b} is an 8-billion parameter model developed by Cisco's Foundation AI team. It is pretrained on a curated corpus of cybersecurity-specific text, including threat intelligence reports, vulnerability databases, incident response documentation, and security standards. Foundation-Sec-8B serves as a domain-adapted base model for applications such as threat detection, vulnerability assessment, security automation, and attack simulation. It enables organizations to build AI-driven security tools that can be deployed locally, reducing dependency on cloud-based AI services while maintaining high performance on security-related tasks.

These generative models complement traditional approaches by providing flexible, context-aware outputs that can be integrated into downstream detection or risk assessment pipelines. Their ability to generate human-readable narratives from structured threat intelligence enhances the interpretability and accessibility of cybersecurity information.

\section{Conclusion and Future Works}
In this work, we introduced \textbf{SecureBERT 2.0}, an enhanced encoder-only language model purpose-built for cybersecurity. Leveraging the ModernBERT architecture with hierarchical encoding and long-context modeling, and pretraining on a multi-modal corpus of over 13 billion text tokens and 53 million code tokens, SecureBERT 2.0 sets new benchmarks across document embedding, named entity recognition, and code vulnerability detection. Our results demonstrate significant gains over prior domain-specific baselines, underscoring the importance of large-scale, security-adapted pretraining for high-stakes applications such as threat intelligence analysis, automated vulnerability detection, and incident triage.

While SecureBERT 2.0 makes substantial progress, there remain several directions for future work:

\begin{enumerate}
    \item \textbf{Scaling Model Capacity} --- Training larger variants of SecureBERT 2.0 with increased parameter counts and extended input length will enable deeper reasoning over multi-file codebases, lengthy incident reports, and complex threat intelligence feeds.
    \item \textbf{Integration with Security Pipelines} --- Embedding SecureBERT 2.0 into operational systems such as SIEMs, SOAR platforms, and secure software development workflows can demonstrate its effectiveness in real-world detection, response, and remediation pipelines.
    \item \textbf{Benchmark Development} --- The cybersecurity community still lacks comprehensive, standardized evaluation datasets. Future efforts should focus on releasing new benchmarks spanning hybrid text--code tasks, vulnerability classification across languages, and longitudinal tracking of emerging threats.
    \item \textbf{Expanding Applications} --- Beyond the tasks explored in this work, SecureBERT 2.0 can be applied to new challenges such as:
    \begin{itemize}
        \item Exploit detection and classification in code repositories,
        \item Log anomaly detection for real-time monitoring,
        \item Policy and compliance mapping, linking natural language security standards to technical enforcement artifacts,
        \item Attack campaign correlation, connecting disparate intelligence reports into coherent threat narratives.
    \end{itemize}
\end{enumerate}
\bibliographystyle{IEEEtran}  
\bibliography{references}  

\begin{thebibliography}{10}
\providecommand{\url}[1]{#1}
\csname url@samestyle\endcsname
\providecommand{\newblock}{\relax}
\providecommand{\bibinfo}[2]{#2}
\providecommand{\BIBentrySTDinterwordspacing}{\spaceskip=0pt\relax}
\providecommand{\BIBentryALTinterwordstretchfactor}{4}
\providecommand{\BIBentryALTinterwordspacing}{\spaceskip=\fontdimen2\font plus
\BIBentryALTinterwordstretchfactor\fontdimen3\font minus \fontdimen4\font\relax}
\providecommand{\BIBforeignlanguage}[2]{{%
\expandafter\ifx\csname l@#1\endcsname\relax
\typeout{** WARNING: IEEEtran.bst: No hyphenation pattern has been}%
\typeout{** loaded for the language `#1'. Using the pattern for}%
\typeout{** the default language instead.}%
\else
\language=\csname l@#1\endcsname
\fi
#2}}
\providecommand{\BIBdecl}{\relax}
\BIBdecl

\bibitem{aghaei2022securebert}
E.~Aghaei, X.~Niu, W.~Shadid, and E.~Al-Shaer, ``Securebert: A domain-specific language model for cybersecurity,'' in \emph{International Conference on Security and Privacy in Communication Systems}.\hskip 1em plus 0.5em minus 0.4em\relax Springer, 2022, pp. 39--56.

\bibitem{warner2024smarter}
B.~Warner, A.~Chaffin, B.~Clavi{\'e}, O.~Weller, O.~Hallstr{\"o}m, S.~Taghadouini, A.~Gallagher, R.~Biswas, F.~Ladhak, T.~Aarsen \emph{et~al.}, ``Smarter, better, faster, longer: A modern bidirectional encoder for fast, memory efficient, and long context finetuning and inference,'' \emph{arXiv preprint arXiv:2412.13663}, 2024.

\bibitem{abdeen2023smet}
B.~Abdeen, E.~Al-Shaer, A.~Singhal, L.~Khan, and K.~Hamlen, ``Smet: Semantic mapping of cve to att\&ck and its application to cybersecurity,'' in \emph{IFIP Annual Conference on Data and Applications Security and Privacy}.\hskip 1em plus 0.5em minus 0.4em\relax Springer, 2023, pp. 243--260.

\bibitem{wang2020minilm}
W.~Wang, F.~Wei, L.~Dong, H.~Bao, N.~Yang, and M.~Zhou, ``Minilm: Deep self-attention distillation for task-agnostic compression of pre-trained transformers,'' 2020.

\bibitem{feng2020codebert}
Z.~Feng, D.~Guo, D.~Tang, N.~Duan, X.~Feng, M.~Gong, L.~Shou, B.~Qin, T.~Liu, D.~Jiang \emph{et~al.}, ``Codebert: A pre-trained model for programming and natural languages,'' \emph{arXiv preprint arXiv:2002.08155}, 2020.

\bibitem{ranade2021CyBERT2}
P.~Ranade, A.~Piplai, A.~Joshi, and T.~Finin, ``Cybert: Contextualized embeddings for the cybersecurity domain,'' in \emph{2021 IEEE International Conference on Big Data (Big Data)}.\hskip 1em plus 0.5em minus 0.4em\relax IEEE, 2021, pp. 3334--3342.

\bibitem{Devlin2019BERTPO}
\BIBentryALTinterwordspacing
J.~Devlin, M.-W. Chang, K.~Lee, and K.~Toutanova, ``Bert: Pre-training of deep bidirectional transformers for language understanding,'' in \emph{North American Chapter of the Association for Computational Linguistics}, 2019. [Online]. Available: \url{https://api.semanticscholar.org/CorpusID:52967399}
\BIBentrySTDinterwordspacing

\bibitem{Liu2019RoBERTaAR}
\BIBentryALTinterwordspacing
Y.~Liu, M.~Ott, N.~Goyal, J.~Du, M.~Joshi, D.~Chen, O.~Levy, M.~Lewis, L.~Zettlemoyer, and V.~Stoyanov, ``Roberta: A robustly optimized bert pretraining approach,'' \emph{ArXiv}, vol. abs/1907.11692, 2019. [Online]. Available: \url{https://api.semanticscholar.org/CorpusID:198953378}
\BIBentrySTDinterwordspacing

\bibitem{paul2022cyberbert}
S.~Paul and S.~Saha, ``Cyberbert: Bert for cyberbullying identification: Bert for cyberbullying identification,'' \emph{Multimedia Systems}, vol.~28, no.~6, pp. 1897--1904, 2022.

\bibitem{hanif2022vulberta}
H.~Hanif and S.~Maffeis, ``Vulberta: Simplified source code pre-training for vulnerability detection,'' in \emph{2022 International joint conference on neural networks (IJCNN)}.\hskip 1em plus 0.5em minus 0.4em\relax IEEE, 2022, pp. 1--8.

\bibitem{bayer2024cysecbert}
M.~Bayer, P.~Kuehn, R.~Shanehsaz, and C.~Reuter, ``Cysecbert: A domain-adapted language model for the cybersecurity domain,'' \emph{ACM Transactions on Privacy and Security}, vol.~27, no.~2, pp. 1--20, 2024.

\bibitem{guo2020graphcodebert}
D.~Guo, S.~Ren, S.~Lu, Z.~Feng, D.~Tang, S.~Liu, L.~Zhou, N.~Duan, A.~Svyatkovskiy, S.~Fu \emph{et~al.}, ``Graphcodebert: Pre-training code representations with data flow,'' \emph{arXiv preprint arXiv:2009.08366}, 2020.

\bibitem{wang2021codet5}
Y.~Wang, W.~Wang, S.~Joty, and S.~C. Hoi, ``Codet5: Identifier-aware unified pre-trained encoder-decoder models for code understanding and generation,'' \emph{arXiv preprint arXiv:2109.00859}, 2021.

\bibitem{ahmad2021unified}
W.~U. Ahmad, S.~Chakraborty, B.~Ray, and K.-W. Chang, ``Unified pre-training for program understanding and generation,'' \emph{arXiv preprint arXiv:2103.06333}, 2021.

\bibitem{zhou2019devign}
Y.~Zhou, S.~Liu, J.~Siow, X.~Du, and Y.~Liu, ``Devign: Effective vulnerability identification by learning comprehensive program semantics via graph neural networks,'' \emph{Advances in neural information processing systems}, vol.~32, 2019.

\bibitem{guo2022unixcoder}
D.~Guo, S.~Lu, N.~Duan, Y.~Wang, M.~Zhou, and J.~Yin, ``Unixcoder: Unified cross-modal pre-training for code representation,'' \emph{arXiv preprint arXiv:2203.03850}, 2022.

\bibitem{xu2022polycoder}
\BIBentryALTinterwordspacing
F.~F. Xu, U.~Alon, G.~Neubig, and V.~J. Hellendoorn, ``A systematic evaluation of large language models of code,'' in \emph{Deep Learning for Code Workshop}, 2022. [Online]. Available: \url{https://openreview.net/forum?id=SLcEnoObJZq}
\BIBentrySTDinterwordspacing

\bibitem{colibri_8b}
{CyberNative-AI}, ``Colibri 8b v0.1,'' \url{https://huggingface.co/CyberNative-AI/Colibri_8b_v0.1}, 2024, accessed: 2025-09-30.

\bibitem{cyberbase_13b}
{CyberNative}, ``Cyberbase 13b,'' \url{https://huggingface.co/CyberNative/CyberBase-13b}, 2024, accessed: 2025-09-30.

\bibitem{zysec_7b}
{ZySec-AI}, ``Zysec-7b,'' \url{https://huggingface.co/ZySec-AI/SecurityLLM}, 2024, accessed: 2025-09-30.

\bibitem{deephat_1_7b}
{DeepHat}, ``Deephat-v1-7b,'' \url{https://huggingface.co/DeepHat/DeepHat-V1-7B}, 2025, accessed: 2025-09-30.

\bibitem{foundation_sec_8b}
{Foundation AI at Cisco}, ``Foundation-sec-8b,'' \url{https://huggingface.co/fdtn-ai/Foundation-Sec-8B}, 2025, accessed: 2025-09-30.

\end{thebibliography}






\end{document}